\documentclass[preprintnumbers,floatfix,aps,prd,a4paper,nofootinbib,superscriptaddress,eqsecnum,notitlepage]{revtex4-1}
\usepackage{amssymb,epsf}
\usepackage{latexsym}
\usepackage{hyperref,natbib,textcase,bm}
\usepackage{amsmath,mathrsfs}
\usepackage{graphicx,epsfig}
\usepackage{latexsym,amssymb}
\usepackage[utf8x]{inputenc}
\usepackage{graphics}
\usepackage[english]{babel}
\usepackage{graphicx}
\usepackage{subfig}
\usepackage{capt-of}
\usepackage{array}
\usepackage{caption}
\usepackage[fit]{truncate}
 \usepackage{fancyhdr}
 \usepackage{comment}
\usepackage{amsfonts}
\usepackage{amssymb,epsfig}
\usepackage{latexsym}
\usepackage{amsmath}
\usepackage{graphicx}

 \graphicspath{{figures}{figures/}{figures/}}

\begin{document}

\title{Observational constraints on G-corrected holographic dark energy  using a Markov chain Monte Carlo method}
\author{Hamzeh Alavirad \footnote{\text{alavirad@kit.edu}}}
\address{Institute for Theoretical Physics, Karlsruhe Institute of Technology, 76128 Karlsruhe, Germany}
\author{Mohammad Malekjani \footnote{\text{malekjani@basu.ac.ir}}}

\address{Department of Physics, Faculty of Science, Bu-Ali Sina University, Hamedan 65178,
Iran\\}

\begin{abstract}
We constrain  holographic dark energy (HDE) 
with   
time varying gravitational coupling constant in the framework 
of the modified Friedmann equations using
cosmological data from type Ia supernovae, baryon acoustic
oscillations, cosmic microwave background radiation and X-ray gas
mass fraction. Applying a Markov Chain Monte Carlo (MCMC) simulation,
we obtain the best fit values of the model and cosmological parameters
within $1\sigma$ confidence level (CL) in a flat
universe as: $\Omega_{\rm
b}h^2=0.0222^{+0.0018}_{-0.0013}$, $\Omega_{\rm
c}h^2=0.1121^{+0.0110}_{-0.0079}$, $\alpha_{\rm G}\equiv
\dot{G}/(HG) =0.1647^{+0.3547}_{-0.2971}$ and the HDE
constant $c=0.9322^{+0.4569}_{-0.5447}$. Using the  best
fit values,  the equation of state of the dark component at the  present time  $w_{\rm d0}$ 
at $1\sigma$ CL can cross the phantom boundary $w=-1$.

\end{abstract}

\maketitle \vspace*{1cm}{\bf keywords:} Cosmology, dark energy,
holographic model, gravitational constant.

\section{Introduction}
The astronomical data from "Type Ia supernova"  \cite{Riess:1998cb, Perlmutter:1998np} 
  indicate that the current universe
 is in an accelerating phase. These observational results have greatly inspirited
 theorists to understand the mechanism of this accelerating
 expansion. 
 In the framework of standard
 cosmology, an exotic energy with negative pressure, the
 so-called dark energy, is attributed to this cosmic acceleration. 
 
Up to now, some theoretical models have been
presented to explain the dynamics of dark energy and cosmic
acceleration of the universe. The simplest but most natural
candidate is the cosmological constant $\Lambda$, with a constant
equation of state (EoS) $w=-1$ \cite{Sahni:1999gb, Peebles:2002gy}.
As we know, the cosmological constant confronts us with two
difficulties: the fine-tuning and cosmic coincidence problems. In
order to solve or alleviate these problems many dynamical dark
energy models with time-varying EoS have been  proposed. The quintessence
\cite{Wetterich:1987fm, Ratra:1987rm}, phantom
\cite{Caldwell:1999ew, Nojiri:2003vn, Nojiri:2003jn}, quintom
\cite{Elizalde:2004mq, Nojiri:2005sx, Anisimov:2005ne}, K-essence
\cite{ArmendarizPicon:2000dh, ArmendarizPicon:2000ah}, tachyon
\cite{Padmanabhan:2002cp, Sen:2002in}, ghost condensate
\cite{ArkaniHamed:2003uy, Piazza:2004df}, agegraphic
\cite{Cai:2007us, Wei:2007ty} and holographic \cite{Witten:2000zk}
are examples of dynamical models. Although many dynamical dark
energy models have been suggested, the nature of dark energy is
still unknown. 

\par Models which are constructed based on fundamental principles are more
preferred as they may exhibit some
underlying features of dark energy. 
Two examples of such kind of dark energy models  are  the agegraphic
\cite{Cai:2007us, Wei:2007ty} and holographic \cite{Hsu:2004ri,
Li:2004rb} models.  In this work we focus on the
holographic dark energy model. The holographic model is built on the
basis of the  holographic principle and some features of quantum gravity
theory \cite{Witten:2000zk}.
 According to the holographic principle, the number of degrees of
 freedom in a bound system should be finite and is related to the
 area of its boundary. In holographic principle,
 a short distance ultra-violet (UV) cut-off is related to the long
distance infra-red (IR) cut-off, due to the limit set by the
formation of a black hole \cite{Horava:2000tb, Thomas:2002pq}. The
total energy of a system with size $L$ should not exceed the mass of a
black hole with the same size, i.e., $L^3\rho_{\rm d}\leq LM_{pl}^2$.
Saturating this inequality, the holographic dark energy density is
obtained as
\begin{equation}\label{1}
\rho_{\rm d}=\frac{3c^{2}}{8\pi GL^{2}}\;,
\end{equation}
where L is the length of the  horizon, $c$ is a numerical constant of
model and $G$ is the gravitational coupling constant.\\
The UV cut-off is related to the vacuum energy and the IR cut-off is
related to the large scale of the universe, such as Hubble
horizon, particle horizon, event horizon, Ricci scalar or the
generalized functions of dimensionless variables as discussed by
\cite{Hsu:2004ri, Li:2004rb,Gao:2007ep, Xu:2009xi}. If we consider
the Hubble length scale for $L$, it leads to wrong equation of state
for dark energy, i.e., $w_d=0$ which can not result the  cosmic
acceleration \cite{Horava:2000tb, Thomas:2002pq}. This problem 
can be cured by considering the interaction between dark matter
and dark energy \cite{Pavon:2005yx, Zimdahl:2007zz}. In the case of
particle horizon, the EoS of dark energy is bigger than $-1/3$,
hence the current accelerated expansion can not be well explained
\cite{Pavon:2005yx, Zimdahl:2007zz}. Holographic dark energy
with event horizon can provide a desired EoS  to describe the
cosmic acceleration \cite{Zhou:2007pz, Sheykhi:2009zv}. 
Nojiri
and Odintsov investigated the HDE model by assuming IR cutoff
depends on the Hubble rate, particle and future horizons \cite{Nojiri:2005pu}. In this
generalized form  the phantom regime can be achieved and also the
coincidence problem is demonstrated. Unification of early phantom
inflation and late time acceleration of the universe is another
feature of this model. 

\par In recent years, the HDE model has been
constrained by various cosmological observations \cite{Huang:2004wt, Enqvist:2004ny,  Enqvist:2004ny, Shen:2004ck, Zhang:2005hs, Kao:2005xp,  Wu:2007fs, Ma:2007pd, Zhang:2007sh, Lu:2009iv, Zhai:2011pp, Zhang:2013mca}.
For example, Huang and Gong  in \cite{Huang:2004wt} obtained the parameter $c$ as
$c=0.21$ by using the SNIa observations.   Enqvist et.al. in
\cite{Enqvist:2004ny} found a connection between the holographic
dark energy and low-$l$ CMB multipoles  by using CMB, LSS and
supernovae data. Zhang et.al. by using the OHD data constrained
the parameter $c$ as $c=0.65^{+0.02}_{-0.03}$ \cite{Zhai:2011pp}.

\par Beside, there are some theoretical and observational evidences
indicating that the gravitational coupling constant $G$ varies with
cosmic time $t$. From the theoretical viewpoint one can be referred to the
works of Dirac\cite{Dirac:1938mt} and Dyson \cite{Dyson:1972, Lannutti:1978dv}. 
In Branse-Dicke theory, the variability of $G$ is also
predicted \cite{Brans:1961sx}. In Kaluza-Klein cosmology, time
varying treatment of $G$ is related to the scalar field appearing in
the metric component corresponding to the 5-th dimension
\cite{Kaluza:1921tu, LorenAguilar:2003kx, Kolb:1985sj, Maeda:1985bq,
Freund:1982pg}. In this case, a
scalar field couples with gravity by  definition of a new parameter.
From observational point of view, the value of the parameter $\dot{G}/G$ (where an overdot represents derivative with respect to the cosmic time $t$) can be
constrained by astrophysical and cosmological observations as well. 
For example data from SNIa observations yields $-10^{-11}{\rm
yr^{-1}}\leq{\dot{G}}/{G}\leq0$ \cite{Gaztanaga:2001fh}.
 The observations of the Binary Pulsar PSR1913
gives $-(1.10\pm1.07)\times10^{-11}{\rm
yr^{-1}}\leq{\dot{G}}/{G}\leq0$ \cite{Damour:1988zz}. The
observational data from the Big Bang nuclei-synthesis results
tighter constraints on this parameter as
 $-3.0\times10^{-13}{\rm yr^{-1}}\leq{\dot{G}}/{G}\leq4.0\times10^{-13}{\rm yr^{-1}}$ \cite{Copi:2003xd}.
 This parameter can be approximated from 
  astro-seismological data from pulsating white
dwarf stars \cite{Benvenuto:2004bs,Biesiada:2003sr} and 
helio-sesmiological \cite{0004-637X-498-2-871} as well. 

\par All mentioned above motivated people to  consider the holographic dark
energy model with time varying gravitational coupling $G$ (G-corrected HDE model) enveloped
by event horizon. In \cite{Jamil:2009sq, Lu:2009iv},  the holographic model with
varying gravitational coupling $G$ was assumed in the  standard Friedmann equations. 
The authors in \cite{Malekjani:2013xsa} considered the G-corrected HDE in the framework of the modified Friedmann equations. 
The holographic model with
varying $G$ in the  standard Friedmann equations has been constrained 
by cosmological data in \cite{Lu:2009iv} where for a flat universe they found $c=0.80^{+0.16}_{-0.13}$ and 
$\alpha_{\rm G}\equiv\dot{G}/(HG) =-0.0016^{+0.0049}_{-0.0019}$. In this paper, by using the cosmological data of 
 Type Ia Supernovae (SNIa), Baryon Acoustic
Oscillations (BAO), Cosmic Microwave Background (CMB) radiation and X-ray gas
mass fraction we will obtain the best fit values of parameters of the G-corrected HDE in the framework of the modified Friedmann equations by applying a Markov Chain Monte Carlo (MCMC) simulation.  
 Based on these best fit values, we obtain the evolution of  EoS and deceleration parameter
$q$ of the $G$-corrected HDE model as well as the  evolution of
energy density parameters. We show that within $1\sigma$ confidence level, this model can
cross the phantom boundary $w=-1$.
 
 \par The paper is organized as
follows: In section \ref{sec:GHDE} the $G$-corrected HDE is
discussed briefly. Then in section \ref{sec:method} the cosmological constraining
method is discussed in detail and the  data fitting results are presented in section \ref{sec:results}. 
The paper is concluded in section \ref{sec:conc}.

\section{G-corrected HDE model in a FRW cosmology}\label{sec:GHDE}
The Hilbert-Einstein action with  time varying gravitational coupling
constant, $G(t)=G_0\phi(t)$, is described as
\begin{equation}
S=\frac{1}{16 \pi G_0}
\int{\sqrt{-g}\left(\frac{R}{\phi(t)}+L_m\right)d^4x}
\end{equation}
where the scalar function $\phi(t)$ is assumed for time dependency
of $G(t)=\phi(t)G_0$, $G_0$ is the bare gravitational coupling constant and $L_m$ is
the lagrangian of the  matter fields. The first modified Friedmann
equation for Robertson-Walker spacetime is obtained as \cite{Malekjani:2013xsa}
\begin{equation}\label{modf}
H^2=\frac{8\pi G(t)}{3} (\rho_{\rm m}+\rho_{\rm d})+H\frac{\dot{G}}{G}
\end{equation}
where an overdot represents the derivative with respect to the cosmic
time t, $H=\dot{a}/a$ and $\rho_{\rm m}$ and $\rho_{\rm d}$ are matter and dark energy densities respectively. We ignore the higher time derivative of $G$ (i.e.,
$\ddot{G}/G$, ...) and also higher powers than one (i.e.,
($\dot{G}/G)^2$, ...), since the value of $\dot{G}/G$ is small
particularly in the late time accelerated universe. The last term on
right hand side of (\ref{modf}) is due to the correction of time
dependency of $G$. Equation (\ref{modf}) can also be obtained
from Branse-Dicke gravity by assuming $w=0$ and $\psi=1/\phi(t)$
in equation (1) of \cite{Banerjee:2007zd} where $w$ is the
Branse-Dicke parameter and $\psi$ is
Branse-Dicke scalar field.\\
Changing the time derivative to a derivative with respect to
$\ln{a}$, equation (\ref{modf}) is expressed as:
\begin{equation}\label{modf1}
H^2(1-\alpha_{\rm G})=\frac{8\pi G(t)}{3} (\rho_{\rm m}+\rho_{\rm d}),
\end{equation}
where $\alpha_{\rm G}={\acute{G}}/{G}$ and the prime represents
derivative with  respect to $\ln a$. Putting $\alpha_{\rm G}=0$ and $G(t)=G_0$,
equation \ref{modf1} reduces to the standard Friedmann equation in flat universe.\\
The energy density of the $G$-corrected HDE model, by assuming the event
horizon IR cut-off
$R_{\rm h}=a\int{\tfrac{dt}{a}}=a\int{\tfrac{H}{\acute{a}}d\acute{a}}$, is
given by
\begin{equation}\label{4}
\rho_{\rm d}=\frac{3c^{2}}{8\pi G(t)R_{\rm h}^{2}}\;.
\end{equation}
Using the definition of dimensionless energy density parameters
$\Omega_{\rm m}=\rho_{\rm m}/\rho_{\rm cr}$ and $\Omega_{\rm d}=\rho_{\rm d}/\rho_{\rm cr}$, where
$\rho_{\rm cr}=3H^2/8\pi G(t)$,
 the modified Friedmann equation (\ref{modf1}) can be rewritten as
\begin{equation}\label{7}
\Omega_{\rm m}+\Omega_{\rm d}=1-\alpha_{\rm G}\;.
\end{equation}
The matter (baryonic and CDM) and dark energy satisfy the following
conservation equation
\begin{subequations}
 \begin{align}
  &\dot{\rho}_m+3H\rho_{\rm m}=0\;,\label{10}\\
  &\dot{\rho}_{\rm d}+3H(1+w_{\rm d})\rho_{\rm d}=0,\label{11}
 \end{align}
\end{subequations}
respectively, where $w_{\rm d}$ is the dark energy EoS. The Hubble parameter in the context of $G$-corrected
HDE model in a flat geometry can be calculated from Eq. (\ref{modf}) as
follows
\begin{equation}\label{17}
H^2(1-\alpha_{\rm G})={H_0}^2[\Omega_{\rm m0} a^{-3}+\Omega_{\rm d0}
a^{-3(1+w_{\rm d})}],
\end{equation}
where $H_0$ is the present value of Hubble parameter and
$\Omega_{\rm m0}$ and $\Omega_{\rm d0}$ are the present values of the
density
parameters of matter (baryonic and CDM) and dark energy respectively.
Taking the time derivative of (\ref{4}) by using conservation
equation (\ref{11}) as well as the relation $\dot{R}_h=1+HR_{\rm h}$
 the equation of state for the G-corrected HDE model can be obtained as
\begin{equation}\label{12}
w_{\rm d}=-\frac{1}{3}-\frac{2}{3}\frac{\sqrt{\Omega_{\rm d}}}{c}+\frac{1}{3}\alpha_{\rm G}\;.
\end{equation}
 The evolutionary
equation of the  dark energy density parameter $\Omega_{\rm d}$ for the  G-corrected HDE model can be
obtained by taking  derivative of
$\Omega_{\rm d}=\tfrac{\rho_{\rm d}}{\rho_{\rm cr}}=\tfrac{c^2}{H^2R_{\rm h}^2}$ with respect
to $\ln{a}$ as follows
\begin{equation}\label{14}
\acute{\Omega_{\rm d}}=-2\Omega_{\rm d}[\frac{c}{HR}+\frac{\dot{H}}{H^2}+1]\;.
\end{equation}
Also, taking the time derivative of the  modified Friedmann equation
(\ref{modf}) yields
\begin{equation}\label{15}
\frac{\dot{H}}{H^2}(1-\frac{1}{2}\alpha_{\rm G})=-\frac{3}{2}(1+w_d\Omega_{\rm d})+2\alpha_{\rm G}\;.
\end{equation}
Inserting (\ref{15}) in (\ref{14}) results
\begin{eqnarray}\label{OL}
\acute{\Omega_{\rm d}}&(1-\alpha_{\rm G}/2)=\Omega_{\rm d}\Big(3(1+w_d\Omega_{\rm d})+\frac{\sqrt{\Omega_{\rm d}}}{c}(2-\alpha_{\rm G})-3\alpha_{\rm G}-2\Big)\;.
\end{eqnarray}
The deceleration parameter $q=-1-\dot{H}/H^2$ for determining the
accelerated phase of the expansion ($q<0$) or decelerated phase
($q>0$) can be obtained for the  $G$-corrected HDE model by using
 (\ref{15}) as
\begin{equation}\label{18}
q(1-\frac{1}{2}\alpha_{\rm G})=\frac{1}{2}(1+3w_d\Omega_{\rm d})-\frac{3}{2}\alpha_{\rm G}\;.
\end{equation}
At early times when the energy density of dark energy tends to zero
and also the correction of $G$ is negligible, one can see that
$q\rightarrow 1/2$, representing deceleration phase in the CDM model.
 In the limiting case of time-independent gravitational
constant G (i.e., $\alpha_{\rm G}=0$) all the above relations reduce to
those obtained for original holographic dark energy (OHDE) model in
\cite{Zhang:2005yz}.

\section{Data Fitting method}\label{sec:method}
The constant $c$ and the quantity $\dot{G}/G$ determine 
evolution of the universe in the G-corrected holographic dark energy
model. Therefor to study the cosmic evolution in 
the G-corrected HDE in the framework of the modified Friedmann equations, it is of great importance to constrain these
parameters by cosmological data.

\par In this section we discuss the method for obtaining the best fit values of the  G-corrected HDE parameters by using the
cosmological data. The fitting method which we use is the  maximum
likelihood method. In this method the total likelihood function
$\mathcal{L}_{\rm tot}=e^{-\chi^2_{\rm tot}/2}$ is maximized by
minimizing $\chi^2_{\rm tot}$. To determine $\chi^2_{\rm tot}$ we
use the following observational data set: cosmic microwave
background radiation (CMB) data from the  seven-year WMAP \cite{Komatsu:2010fb}, type
Ia supernova (SNIa) data from 557 Union2 \cite{Union2} , baryon acoustic oscillation (BAO) data from
SDSS DR7 \cite{Percival:2009xn}, and cluster X-ray gas mass fraction
data which is measured by Chandra X-ray telescope observations \cite{Allen:2007ue}. Therefor $\chi^2_{\rm tot}$ is given by the
relation
\begin{equation}
\chi^2_{\rm tot}=\chi^2_{\rm SNIa}+\chi^2_{\rm CMB}+\chi^{2}_{\rm
BAO}+\chi^2_{\rm gas}\;.\label{eq:totchi1}
\end{equation}
In following we discuss each $\chi^2$ in detail. 

 The data for SNIa are 557 Union2 data \cite{Union2}. In this case $\chi^2_{\rm SNIa}$ is obtained
by comparing the theoretical distance modulus $\mu_{\rm th}(z)$ with
the observed one $\mu_{\rm ob}(z)$
\begin{equation}
 \chi^2_{\rm SNIa}=\sum_{\rm i}\frac{[\mu_{\rm th}(z_i)-\mu_{\rm obs}(z_i)]^2}{\sigma_{\rm i}^2}\;,\label{eq:x2:SNIa}
\end{equation}
with \begin{equation}
 \mu_{\rm th}(z)=5\log_{10}[D_{\rm L}(z)]+\mu_{0}\; ,\label{muth}
\end{equation}
where $\mu_{0}=5\log_{10}(cH_{0}^{-1}/Mpc)+25$ and the observational modulus distance of SNIa, $\mu_{\rm obs}(z_i)$,
at redshift $z_i$ is given by
\begin{equation}
 \mu_{\rm obs}(z_i)=m_{\rm obs}(z_i)-M,\label{muobs}
\end{equation}
where $m$ and $M$ are apparent and absolute magnitudes of SNIa respectively. The Hubble-free luminosity distance $D_{\rm L}$ is given by
\begin{equation}
 d_{\rm L}(z)=\frac{H_0(1+z)}{\sqrt{|\Omega_{\rm k}|}}sinn[\sqrt{|\Omega_{\rm k}|}\int_{0}^{z}\frac{dz'}{H(z')}]\;,
\end{equation}
where $sinn(\sqrt{|\Omega_{\rm k}|}x)$ represents respectively $sin(\sqrt{|\Omega_{\rm k}|}x)$, $\sqrt{|\Omega_{\rm k}|}$ and $sinh(\sqrt{|\Omega_{\rm k}|})$
for $\Omega_{\rm k}<0$, $\Omega_{\rm k}=0$ and $\Omega_{\rm k}>0$.
Eq. (\ref{eq:x2:SNIa}) can be written \cite{Nesseris:2007pa}
\begin{equation}
 \chi^2_{SNIa}=A+2B\mu_0+C\mu_0^2\;,
\end{equation}
where
\begin{subequations}
 \begin{align}
  A&=\sum_{i}\frac{[\mu_{\rm th}(z_i;\mu_0=0)-\mu_{\rm obs}(z_i)]^2}{\sigma_{\rm i}^2}\label{eq:A}\\
  B&=\sum_{i}\frac{\mu_{\rm th}(z_i;\mu_0=0)-\mu_{\rm obs}(z_i)}{\sigma_{\rm i}^2}\label{eq:B}\\
  C&=\sum_{i}\frac{1}{\sigma_{\rm i}^2}
 \end{align}
\end{subequations}
where $\mu_0=42.384-5\log_{10}h$. The minimum of  eq. (\ref{eq:x2:SNIa}) can be written as 
\begin{equation}
 \chi^{2}_{\rm SNIa,min}=A-B^2/C\;.
\end{equation}
The goodness of fit between the theoretical model and data is
expressed by $\chi^{2}_{\rm SNIa,min}$.

For the CMB data,  we use the data points $(R, l_{\rm a}, z_{\ast})$ from seven-year WMAP \cite{Komatsu:2010fb}.
The data points parameters are as follows: $R$ is the scaled
distance to recombination $R=\sqrt{\Omega_{\rm
m0}/c}\int_{0}^{z_{\ast}}dz/E(z)$, where $E(z)\equiv H(z)/H_0$ and 
$z_{\ast}$ is recombination redshift \cite{Hu:1995en}. The angular
scale of the sound horizon at recombination is given by $l_{\rm A}$
\cite{Bond:1997wr}
\begin{equation}
 l_{\rm A}=\frac{\pi r(z_{\ast})}{r_{\rm s}(z_{\ast})}\;\label{lCMB}\;,
\end{equation}
where $r(z)$ is the comoving distance $
r(z)=c/H_{0}\int_{0}^{z}dz^{\prime}/E(z^{\prime})$ and  the comoving
sound horizon distance at the recombination $r_{\rm s}(z_{\ast})$ is
given by
\begin{equation}
 r_{\rm s}(z_{\ast})=\int_{0}^{a(z_{\ast})}\frac{c_{\rm s}(a)}{a^2H(a)}da\;,\label{shdis}
\end{equation}
where the sound speed $c_{\rm s}(a)$ is defined by
\begin{equation}
 c_{\rm s}(a)=\left[3(1+\frac{3\Omega_{\rm b0}}{4\Omega{\gamma0}}a)\right]^{-1/2}\;,\label{soundspeed}
\end{equation}
Seven-year WMAP observations give $\Omega_{\rm\gamma0}=2.469\times10^{-5}h^{-2}$ and $\Omega_{\rm b0}=0.02260\pm0.00053\times10^{-5}h^{-2}$ \cite{Komatsu:2010fb}.

\par The recombination redshift $z_{\ast}$ is obtained  using the fitting function proposed by Hu and Sugiyama \cite{Hu:1995en}
\begin{equation}
 z_{\ast}=1048[1+0.00124(\Omega_{\rm b0}h^2)^{-0.738}][1+g_{\rm 1}(\Omega_{\rm m0}h^2)^{g_{2}}]\;,\label{redshift}
\end{equation}
where $g_1=(0.0783(\Omega_{\rm b0}h^2)^{-0.238})/(1+39.5(\Omega_{\rm
b0}h^2)^{0.763})$ and $g_2=({0.560})/(1+21.1(\Omega_{\rm
b0}h^2)^{1.81})$. Then one can define  $\chi^2_{\rm CMB}$  as
$\chi^2_{\rm CMB}=X^TC^{-1}_{\rm CMB}X$, with \cite{Komatsu:2010fb}
\begin{subequations}
\begin{align}
 X&=\begin{pmatrix} l_{\rm A}-302.09 \\R-1.725\\z_{\ast}-1091.3 \end{pmatrix},\label{CCMB}\\
 C^{-1}_{\rm CMB}&=\begin{pmatrix} 2.305 & 29.698 &-1.333 \\293689 & 6825.270 & -113.180\\-1.333 & -113.180 & 3.414 \end{pmatrix},\label{invcovCMB}
 \end{align}
\end{subequations}
where $C^{-1}_{\rm CMB}$ is the inverse covariant matrix.

 The data from Sloan Digital Sky Survey
(SDSS) Data Release 7 (DR7) \cite{Percival:2009xn} is used for the baryon
acoustic oscillations (BAO) data. One can define $\chi^2_{\rm BAO}$ by
$\chi^2_{\rm BAO}=Y^TC^{-1}_{\rm BAO}Y$, where
\begin{subequations}
\begin{align}
Y&= \begin{pmatrix} d_{0.2}-0.1905\\d_{0.35}-0.1097\end{pmatrix}\; ,\label{YBAO}\\
 C^{-1}_{\rm BAO}&=\begin{pmatrix}30124 & -17227\\-17227 & 86977\end{pmatrix}\; .\label{CBAO}
 \end{align}
\end{subequations}
The data points $d_{z_{i}}$ is defined as $d_{\rm z_{i}}\equiv r_{\rm
s}(z_{\rm d})/D_{\rm V}(z_{i})$, where $r_{\rm s}(z_{\rm d})$ is the
comoving sound
 horizon distance at the drag epoch (where baryons were released from photons) and $D_{\rm V}$ is given by \cite{Eisenstein:2005su}
\begin{equation}
D_{\rm V}(z)\equiv\left[
\left(\int_{0}^{z}\frac{dz^{\prime}}{H(z^{\prime})}\right)^2\frac{cz}{H(z)}\right]^{1/3}\;.\label{DBAO}
\end{equation}
The drag redshift is given by the fitting formula
\cite{Eisenstein:1997ik}
\begin{equation}
 z_{\rm d}=\frac{1291(\Omega_{\rm m0}h^2)^{0.251}}{1+0.659(\Omega_{\rm m0}h^2)^{0.828}}\left[1+b_{1}(\Omega_{\rm b0}h^2)^{b_{2}}\right]\; ,\label{dragredshift}
\end{equation}
where $b_{1}=0.313(\Omega_{\rm m0}h^2)^{-0.419}[1+0.607(\Omega_{\rm
m0}h^2)^{0.607}]$ and $b_{2}=0.238(\Omega_{\rm m0}h^2)^{0.223}$.\\

 The final data we use is X-ray gas mass fraction data from the Chandra X-ray observations \cite{Allen:2007ue}.
In this case we use the definition $\chi^2_{\rm gas}$ 
\begin{eqnarray}
\chi^2_{\rm gas}&=&\sum_i^N\frac{[f_{\rm gas}^{\Lambda
CDM}(z_i)-f_{\rm gas}(z_i)]^2}{\sigma_{\rm f_{\rm
gas}}^2(z_i)}+\frac{(s_{0}-0.16)^{2}}{0.0016^{2}}
\\ \nonumber &+&\frac{(K-1.0)^{2}}{0.01^{2}}+\frac{(\eta-0.214)^{2}}{0.022^{2}}\;,\label{chifgas}
\end{eqnarray}
where $s_0=(0.16\pm0.05)h_{70}^{0.5}$, $K=1.0\pm0.1$ and
$\eta=0.214\pm0.022$ \cite{Allen:2007ue}. The details for  the from of
the mass gas fractions $f_{\rm gas}(z)$ and $f_{\rm gas}^{\rm
\Lambda CDM}$ is discussed in \cite{Allen:2007ue}.

\section{Data fitting results}\label{sec:results}

\par
\begin{figure}[t]
\centering
\begin{tabular}{ccccc}
\epsfig{file=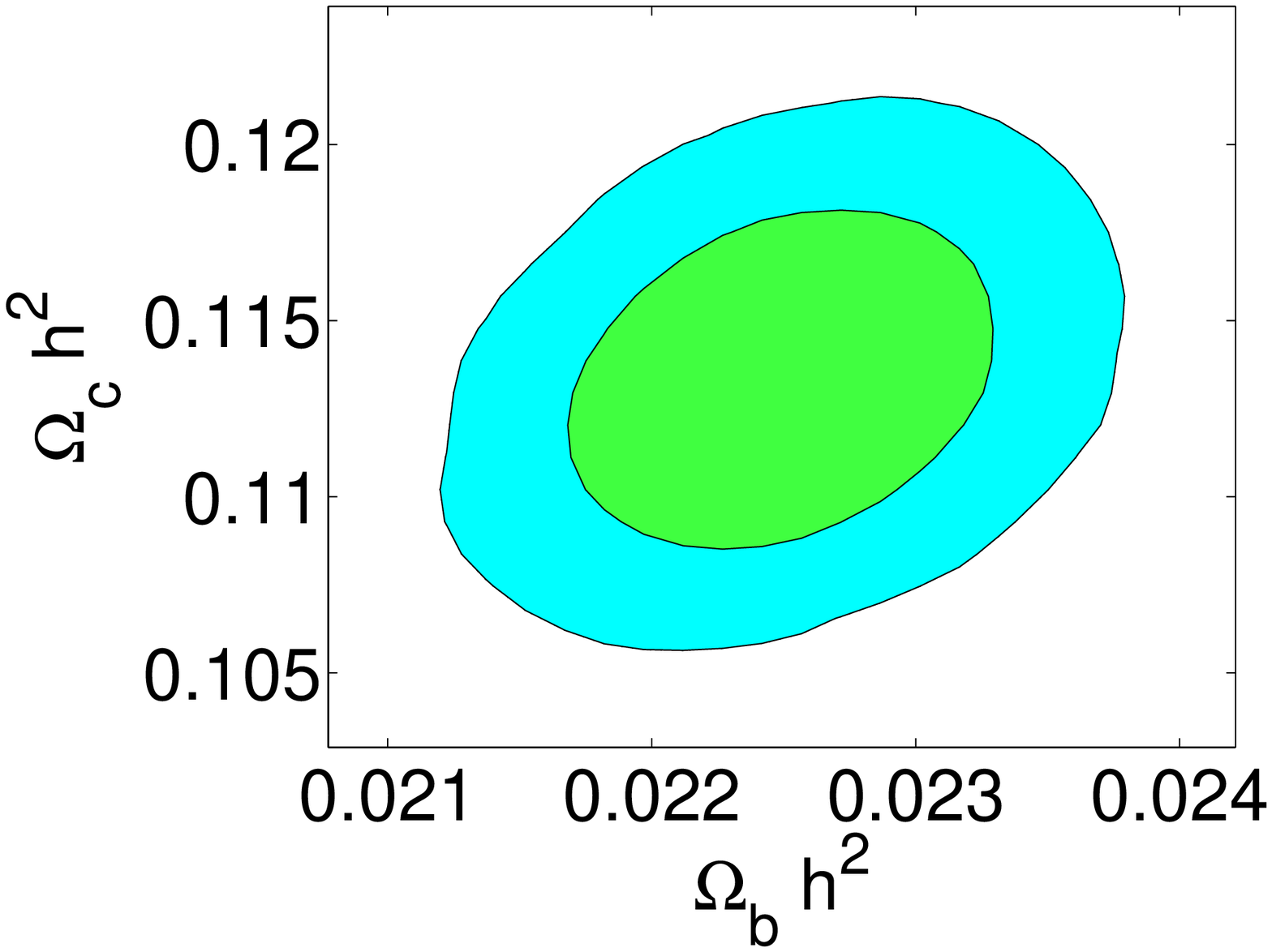,width=0.22\linewidth} & & & &  \\
\epsfig{file=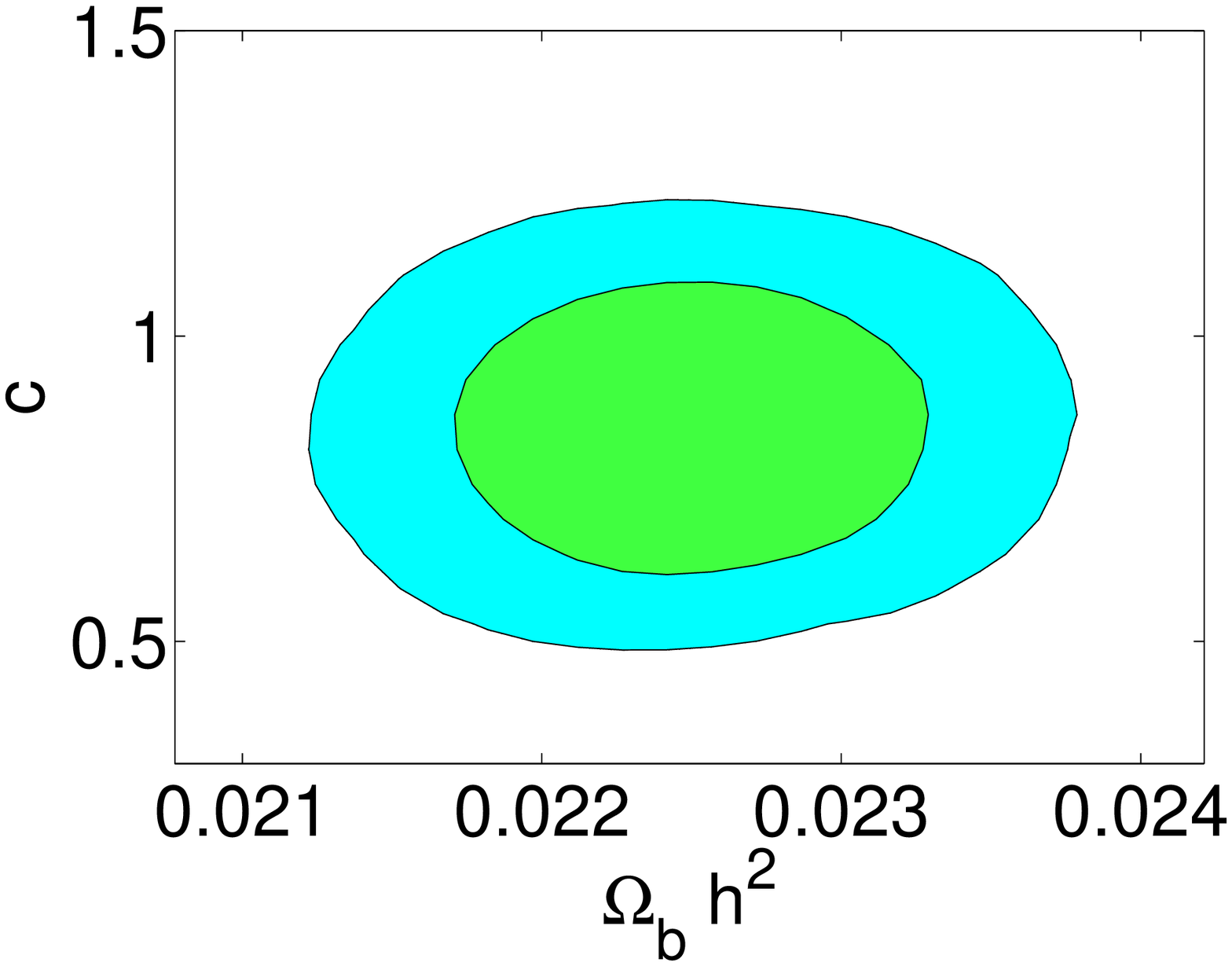,width=0.2\linewidth} &\epsfig{file=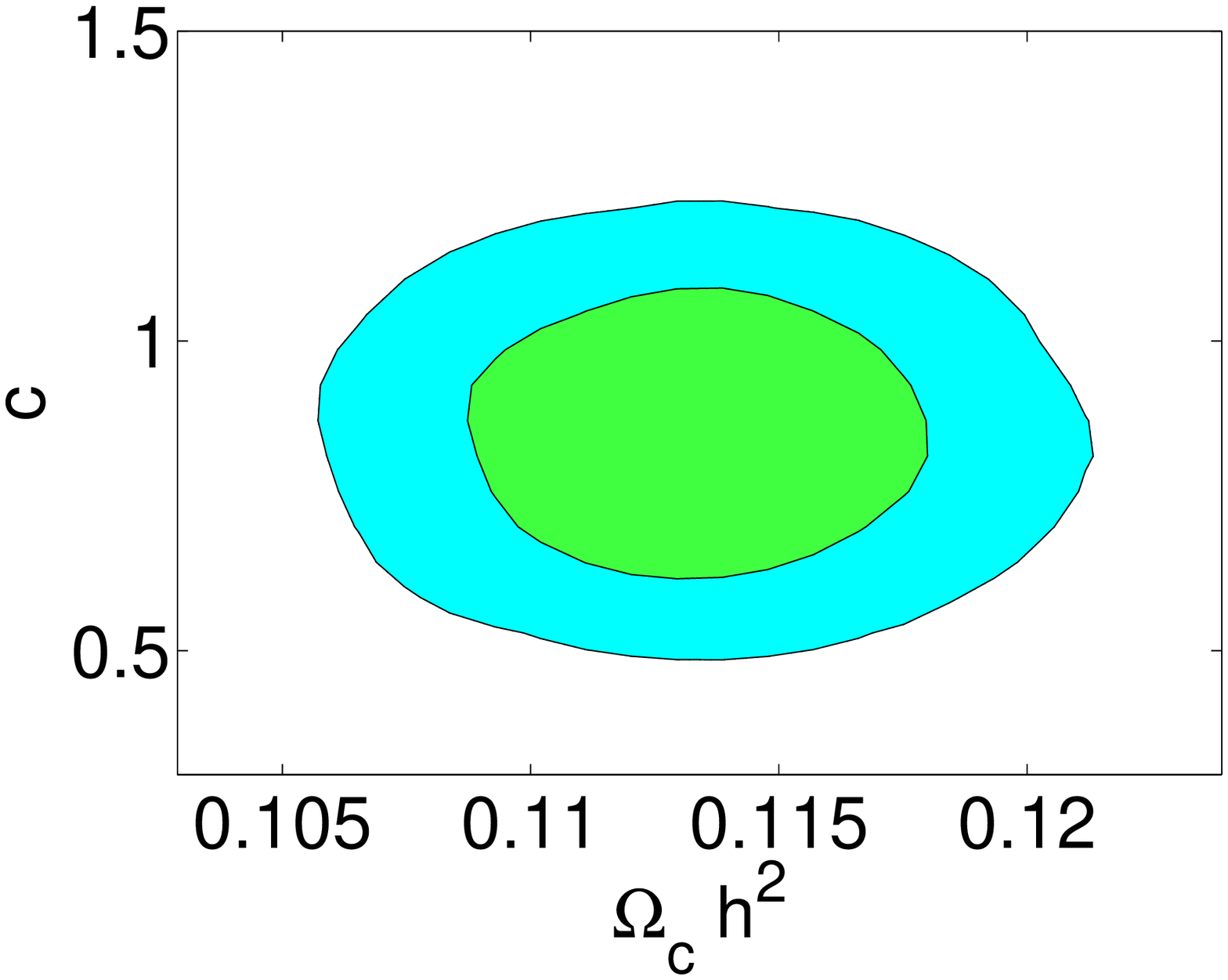,width=0.2\linewidth} \\
\epsfig{file=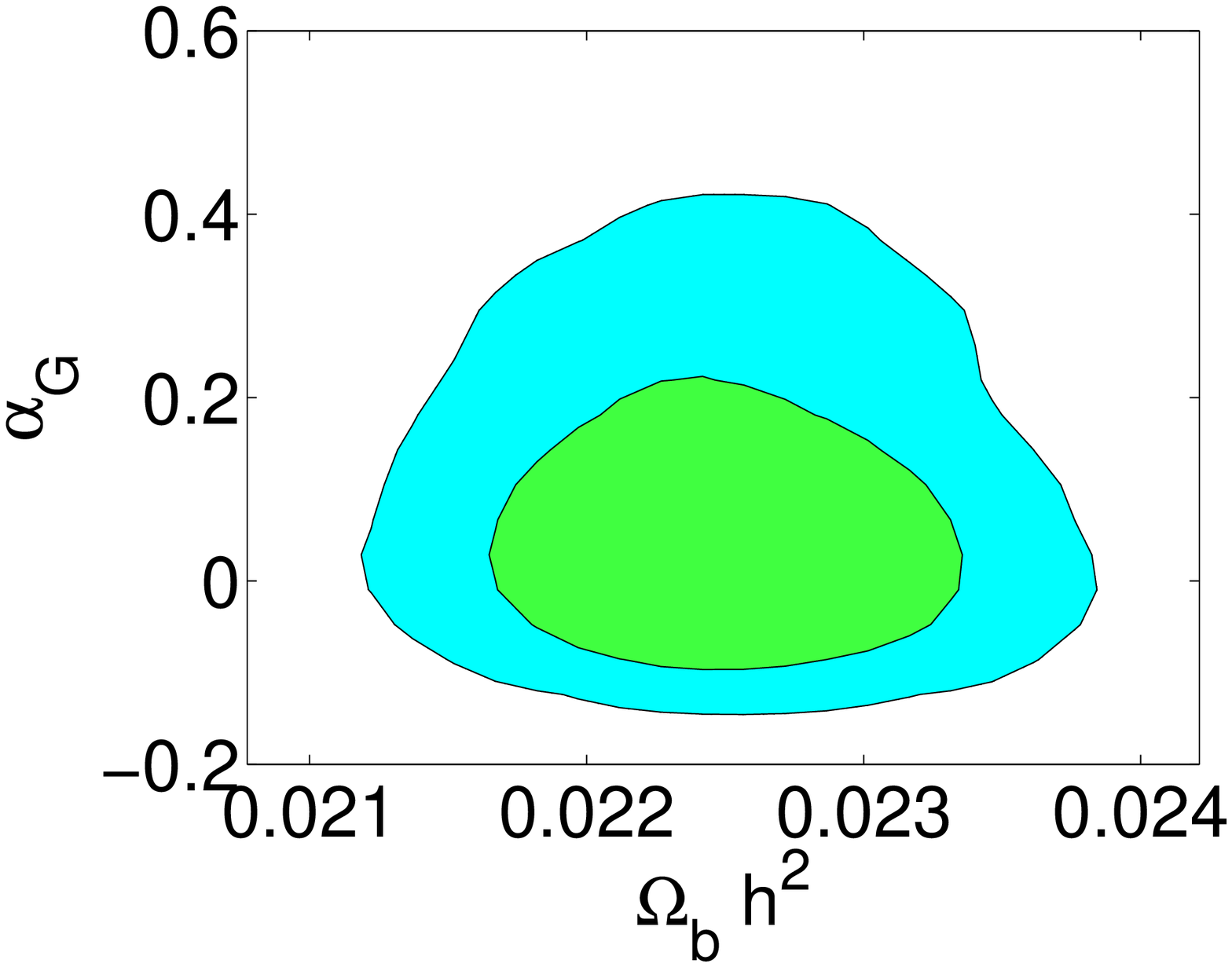,width=0.2\linewidth} &\epsfig{file=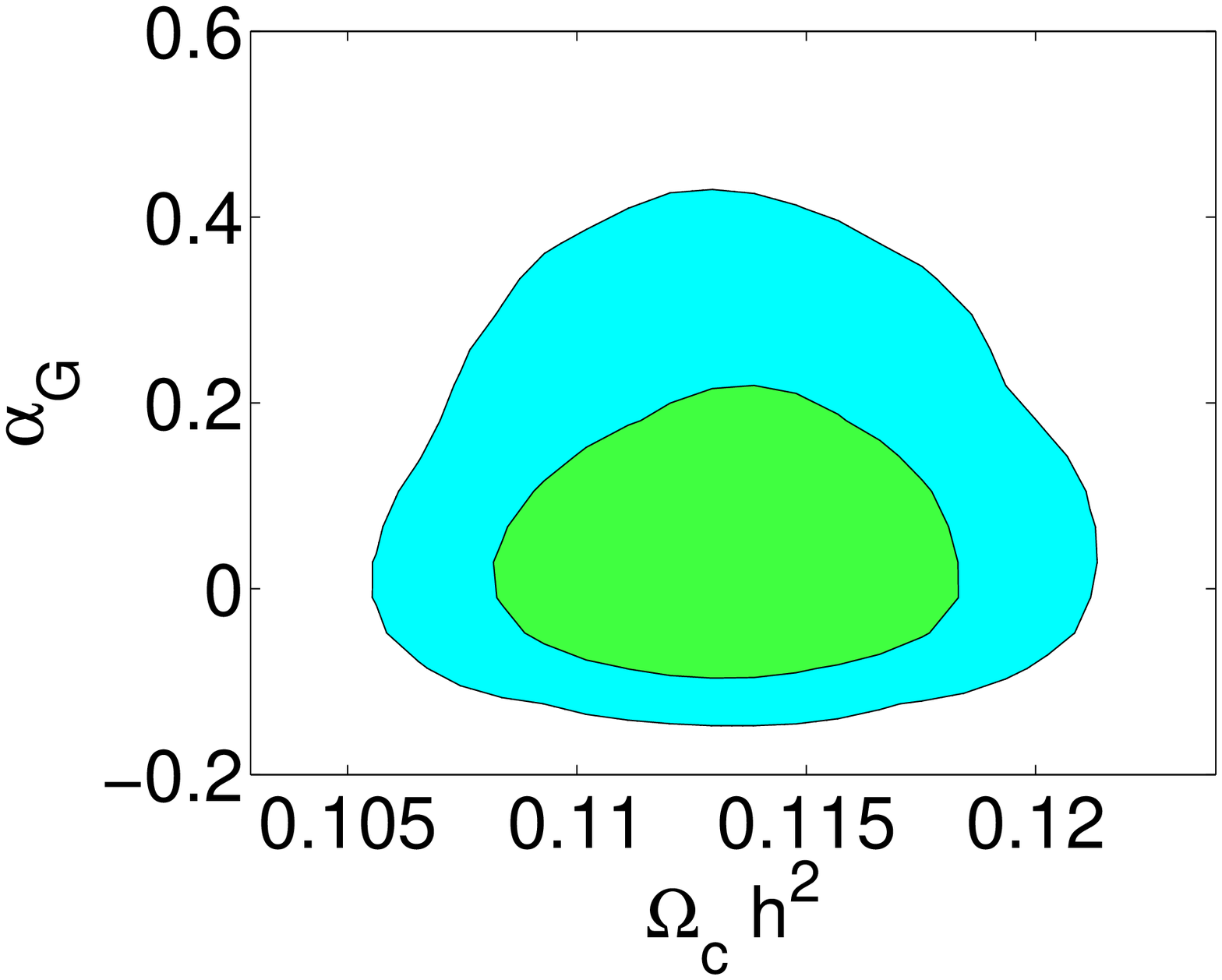,width=0.2\linewidth} & \epsfig{file=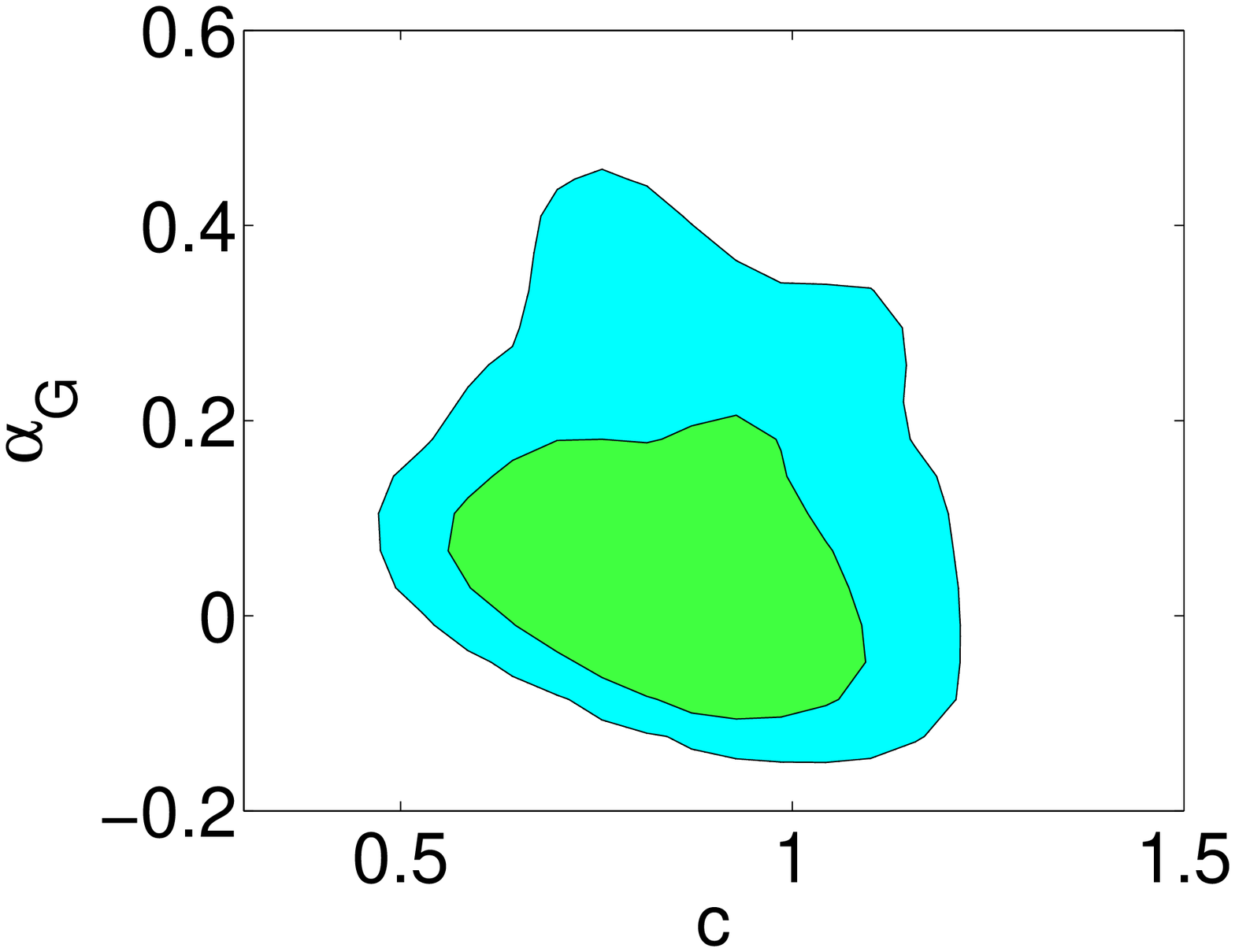,width=0.2\linewidth}\\
\epsfig{file=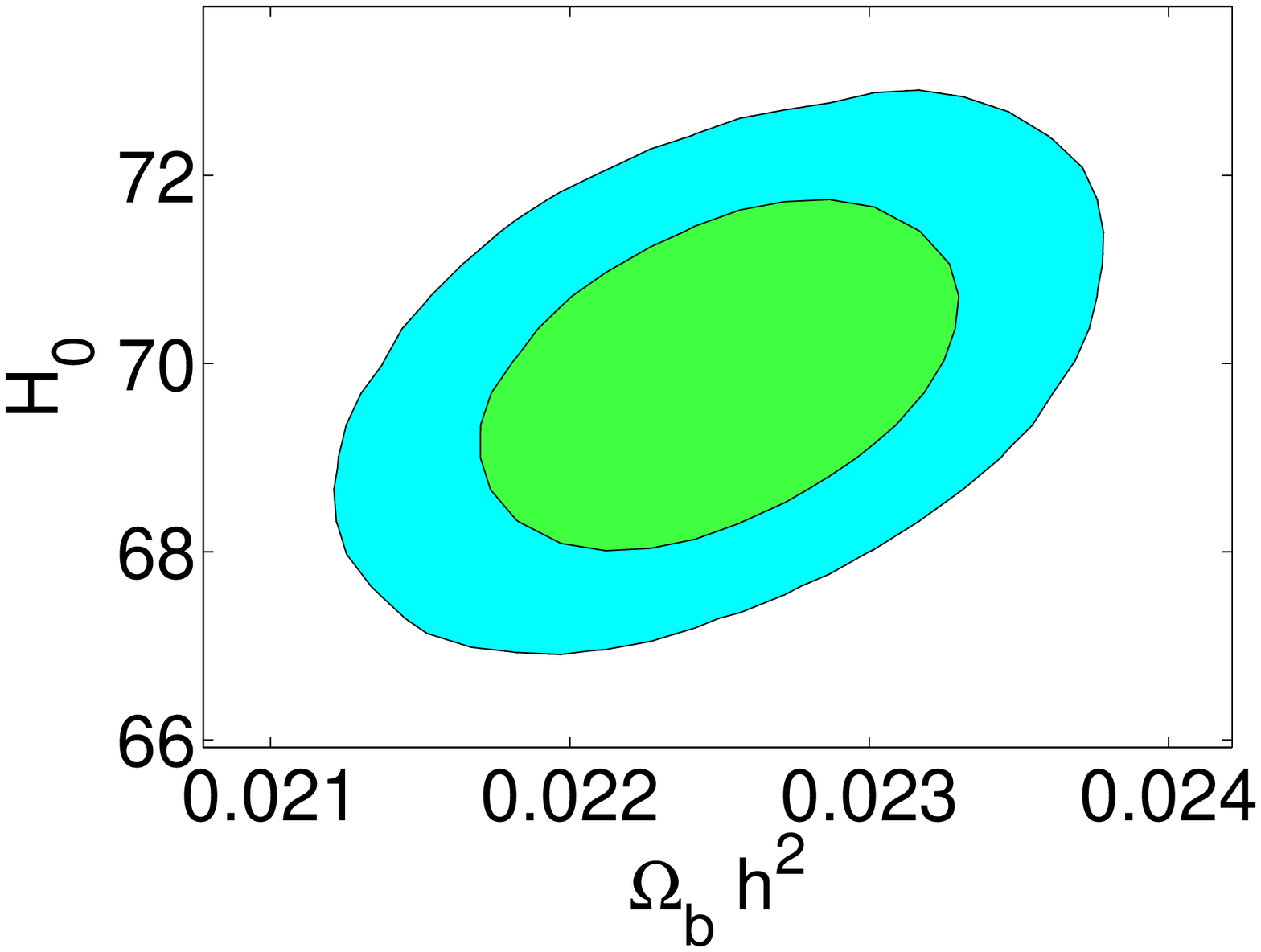,width=0.2\linewidth} &\epsfig{file=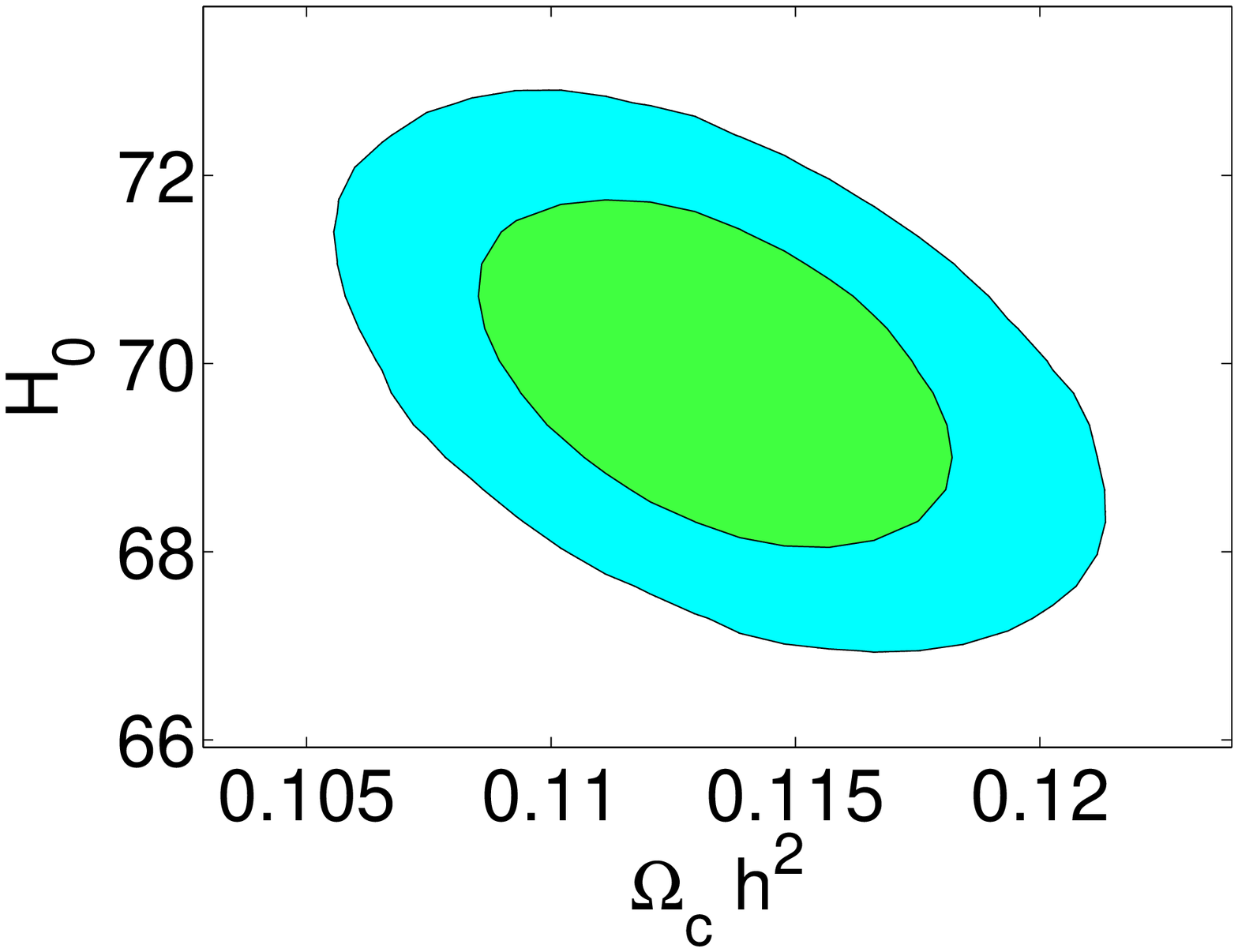,width=0.2\linewidth} & \epsfig{file=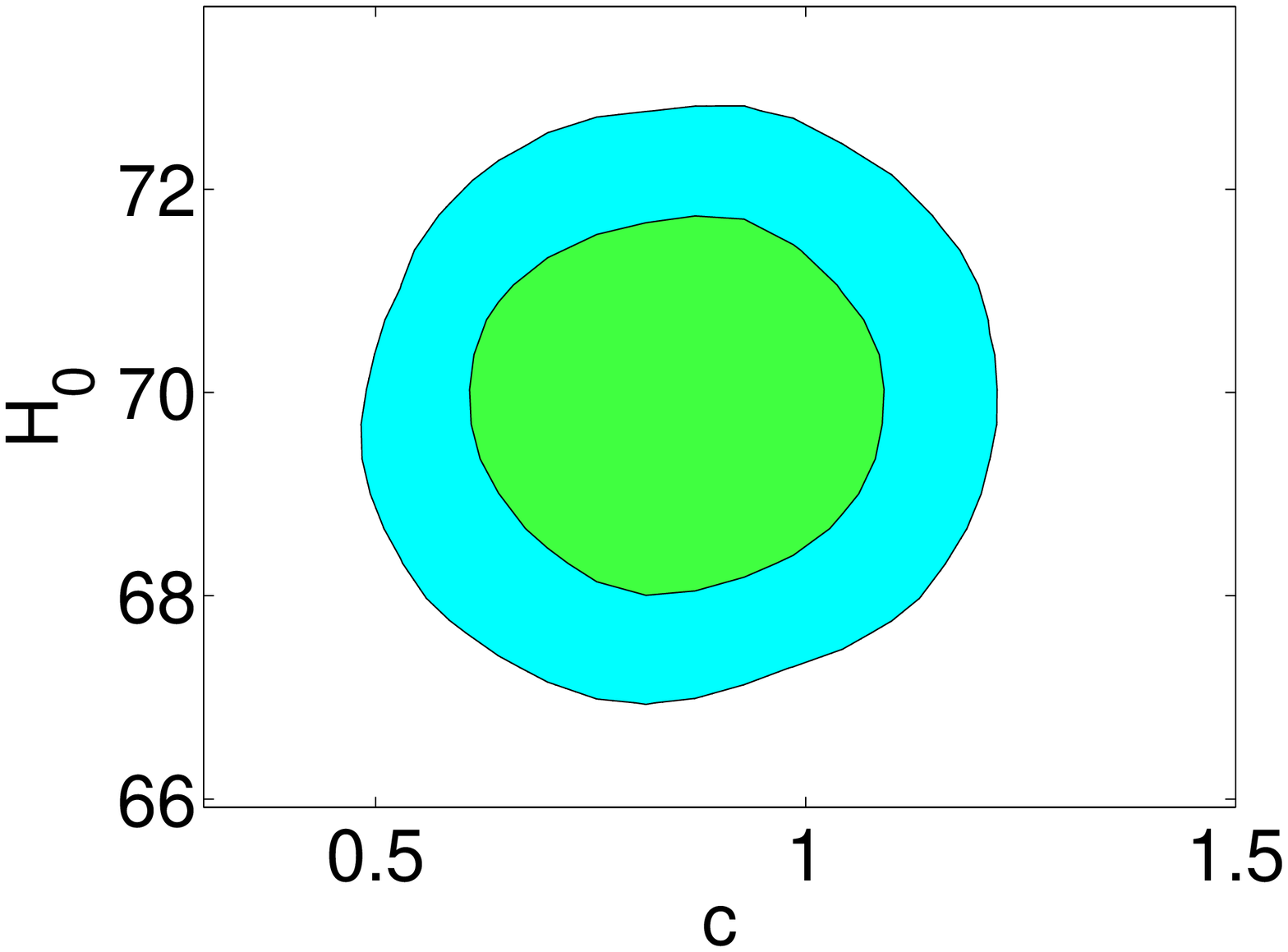,width=0.2\linewidth}& \epsfig{file=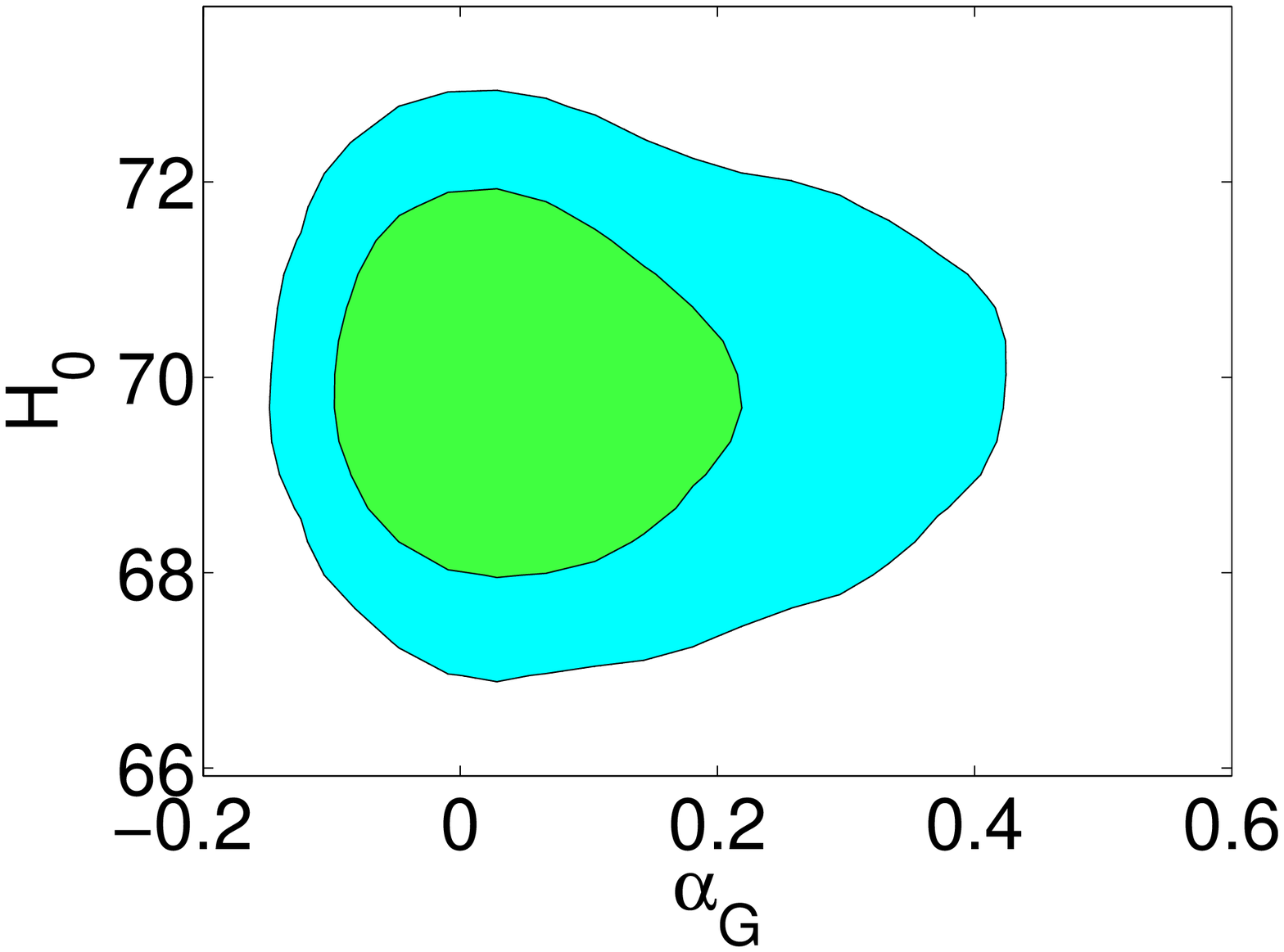,width=0.2\linewidth}\\
\end{tabular}
\caption[2-dimensional contours with $1\sigma$ and $2\sigma$
regions.] {2-dimensional constraint of the cosmological and model
parameters contours in the G-corrected HDE model with $1\sigma$ and
$2\sigma$ regions. To produce these plots, SNIa+CMB+BAO+X-ray gas
mass fraction data sets together with the BBN constraints have been
used.}\label{fig:MCMC}
\end{figure}

 Finally we apply a Markov Chain Monte Carlo simulation on the G-corrected HDE model
by modifying the publically available CosmoMC code \cite{cosmomc}.
The parameter space is chosen as $(\Omega_{\rm b}h^2, \Omega_{\rm
c}h^2, \alpha_{\rm G}, c)$ with the priors $\Omega_{\rm b}h^2=[0.005,\; 0.1]$,
$\Omega_{\rm c}h^2=[0.01,\; 0.99]$, $\alpha_{\rm G}=[-1,\;+1]$ and $c=[0,\;2]$.
We also consider the derived parameters $(\Omega_{d},\; H_{0},\; {\rm age})$ as well. The results of the best fit values
are presented in table \ref{tab:MCMC}. In addition figure
\ref{fig:MCMC} shows the 2-dimensional constraints of the
cosmological parameters contours with $1\sigma$ and $2\sigma$
confidence levels.

\begin{table}[t]
\begin{center}
    \begin{tabular}{|c|c|c|}
        \hline
        Parameter     & Best Fit Value & $\Lambda$CDM  \\ \hline\hline
        $\Omega_{\rm b}h^2$ & ~   $0.0222^{+0.0018+0.0021}_{-0.0013-0.0016}$ & $0.02214\pm0.00024$  \\ \hline
        $\Omega_{\rm c}h^2 $& ~    $0.1121^{+0.0110+0.0130}_{-0.0079-0.0096}$  & $0.1187\pm0.0017 $   \\ \hline
        $\Omega_{\rm d}$  & ~      $0.7246^{+0.0342+0.0418}_{-0.0485-0.0606}$ & $0.692\pm0.010 $\\ \hline
    $c$         & ~       $0.9322^{+0.4569}_{-0.5447}$ & $\ldots$    \\ \hline
    $\alpha_{\rm G}$    &~ $0.1647^{+0.3547+0.3576}_{-0.2971-0.2978}$    & $\ldots$ \\ \hline
        $H_{0}$         & ~     $69.8809^{+3.5339+4.1638}_{-3.4423-4.4567}$   & $67.80\pm0.77$ \\ \hline
         Age (Gyr)       & ~      $13.8094^{+0.2801+0.3776}_{-0.3618-0.4392}$  &~ $13.798\pm0.45$\\ \hline
    \end{tabular}
\caption[The best fit values of the cosmological and model
parameters in the G-corrected HDE model with $1\sigma$ and $2\sigma$
regions. Here CMB, SNIa BAO and X-ray gas mass fraction data sets are used.] {The
best fit values of the cosmological and model parameters in the
G-corrected HDE model with $1\sigma$ and $2\sigma$ regions. Here the
CMB, SNIa, BAO and X-ray gas mass fraction data together with the BBN constraints have been used. For comparison,
the results for the $\Lambda$CDM model from the Planck data are
presented in the third column \cite{Ade:2013zuv}.} \label{tab:MCMC}
\end{center}
\end{table}

\par From table \ref{tab:MCMC} one can see that all main cosmological parameters ($\Omega_{\rm b}h^2$, $\Omega_{\rm c}h^2$,
$\Omega_{\rm d}$, $H_0$, age) are in agreement with the results of the
$\Lambda$CDM model \cite{Ade:2013zuv} as one can see in the third
column. 
The best fit value of the parameter $c$ i.e. 
$c=0.9322^{+0.4569}_{-0.5447}$  is also compatible with other works
such as $c=0.91^{+0.21}_{-0.18}$ in \cite{Zhang:2007sh},
$c=0.84^{+0.14}_{-0.12}$ in \cite{Zhang:2013mca} and
$c=0.68^{+0.03}_{-0.02}$ in \cite{Zhai:2011pp}. Then by using the
best fit values of parameters $\alpha_{\rm G}$ and $H_0$ one can obtain
approximately the best fit value of quantity
$\dot{G}/G=+1.14\times10^{-11}{\rm yr^{-1}}$. This results is in
agreement with the results of other constraining works. For example
the astroseismological data obtained from pulsating white dwarf
stars result $-2.5\times10^{-10}{\rm
yr^{-1}}\leq{\dot{G}}/{G}\leq+4.5\times10^{-10}{\rm yr^{-1}}$
\cite{Benvenuto:2004bs} and observations of the pulsating white
dwarf G117-B15A suggest ${\dot{G}}/{G}\leq+4.1\times10^{-11}{\rm
yr^{-1}}$  \cite{Biesiada:2003sr}. Therefore these two best values
offer a self-consistency for our analysis. Lu et.al. in \cite{Lu:2009iv}
constrained HDE with varying gravitational coupling constant by using 
SNIa, CMB, BAO and OHD (Observational Hubble Data) data in the standard Friedmann equations framework. 
They found  the best fit values: $c=0.80^{+0.16}_{-0.13}$ and 
$\alpha_{\rm G}=-0.0016^{+0.0049}_{-0.0019}$. 
Our results in $1\sigma$ CL are comparable with the Lu et. al. results as well. 

\begin{figure}[t]
\includegraphics[width=0.4\linewidth]{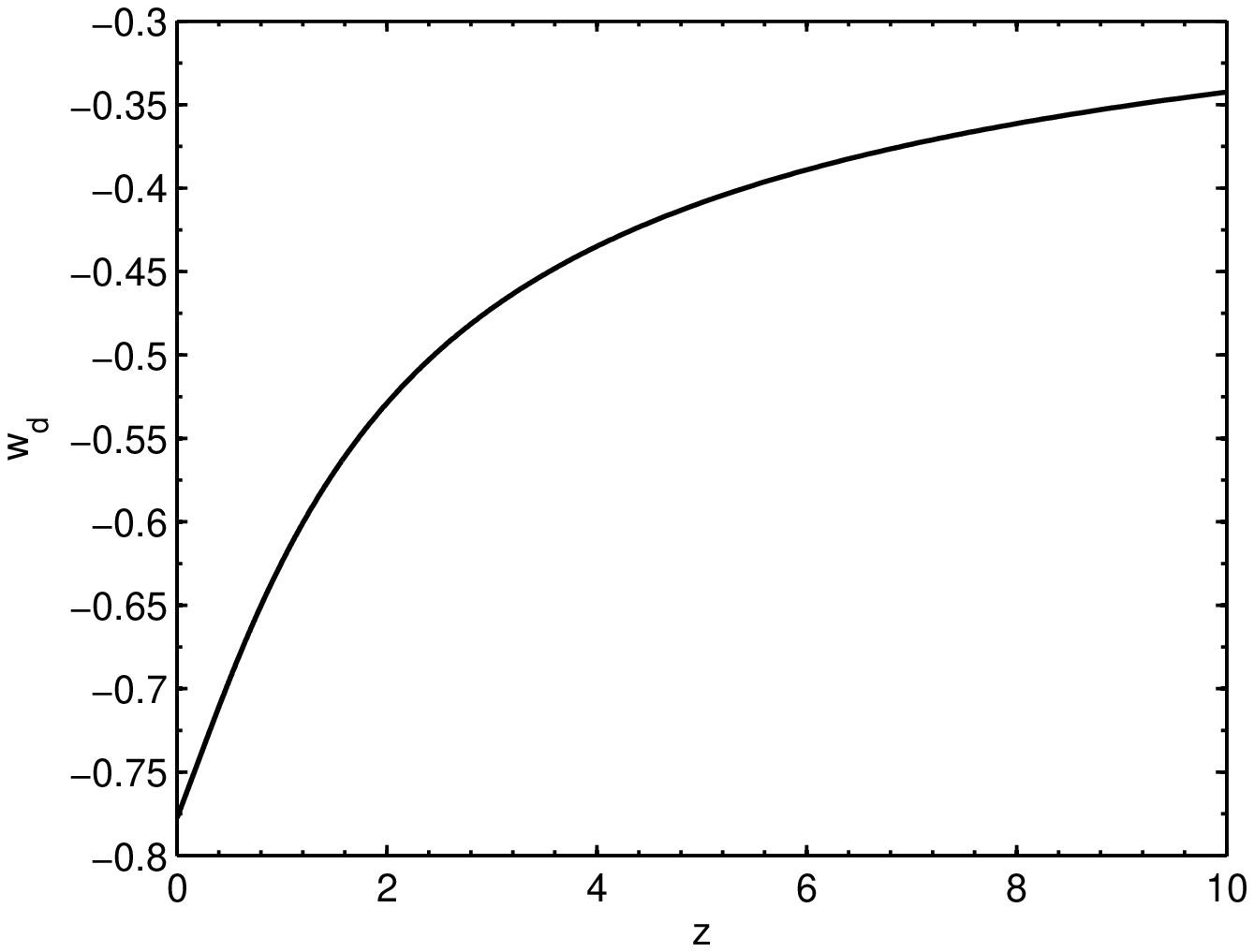}
\includegraphics[width=0.4\linewidth]{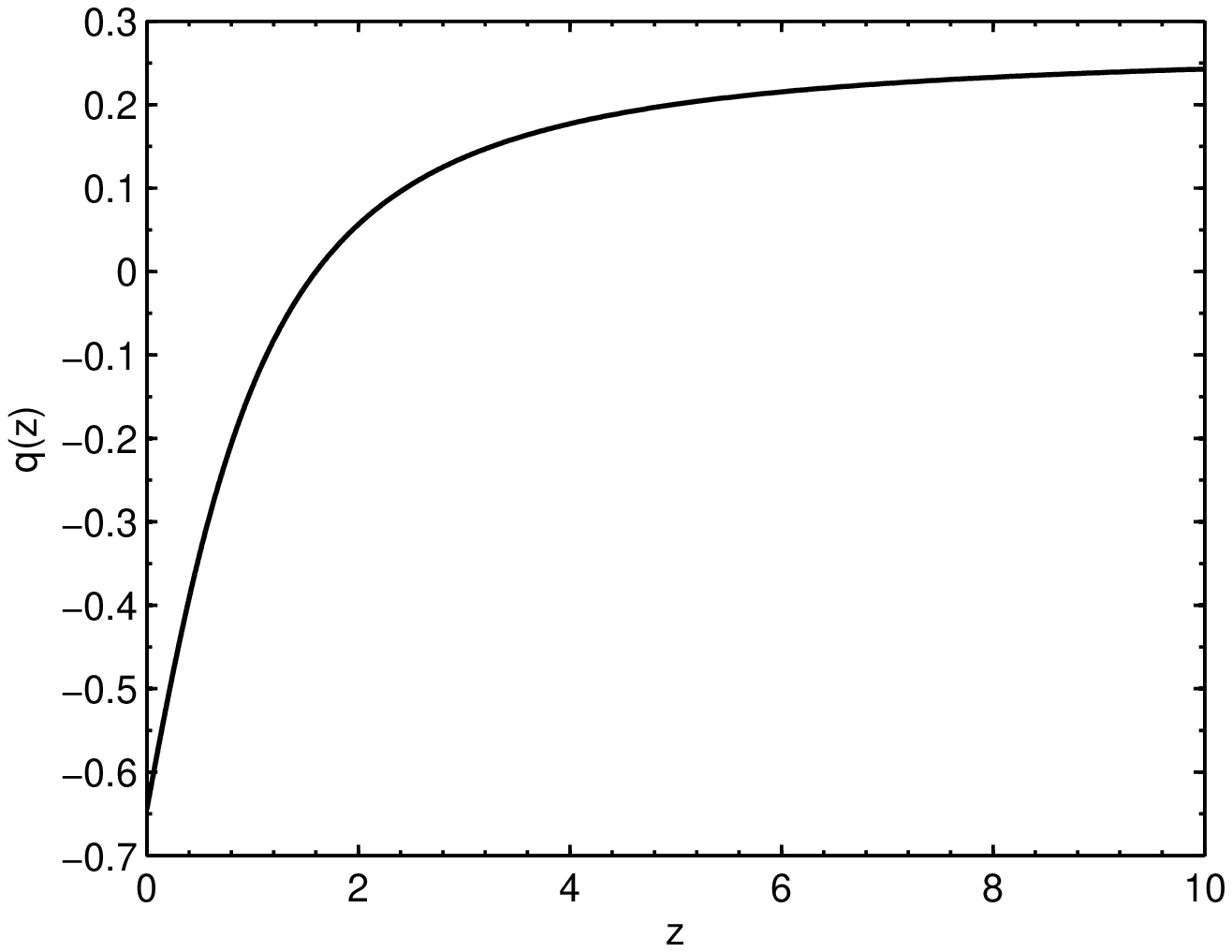}
\includegraphics[width=0.4\linewidth]{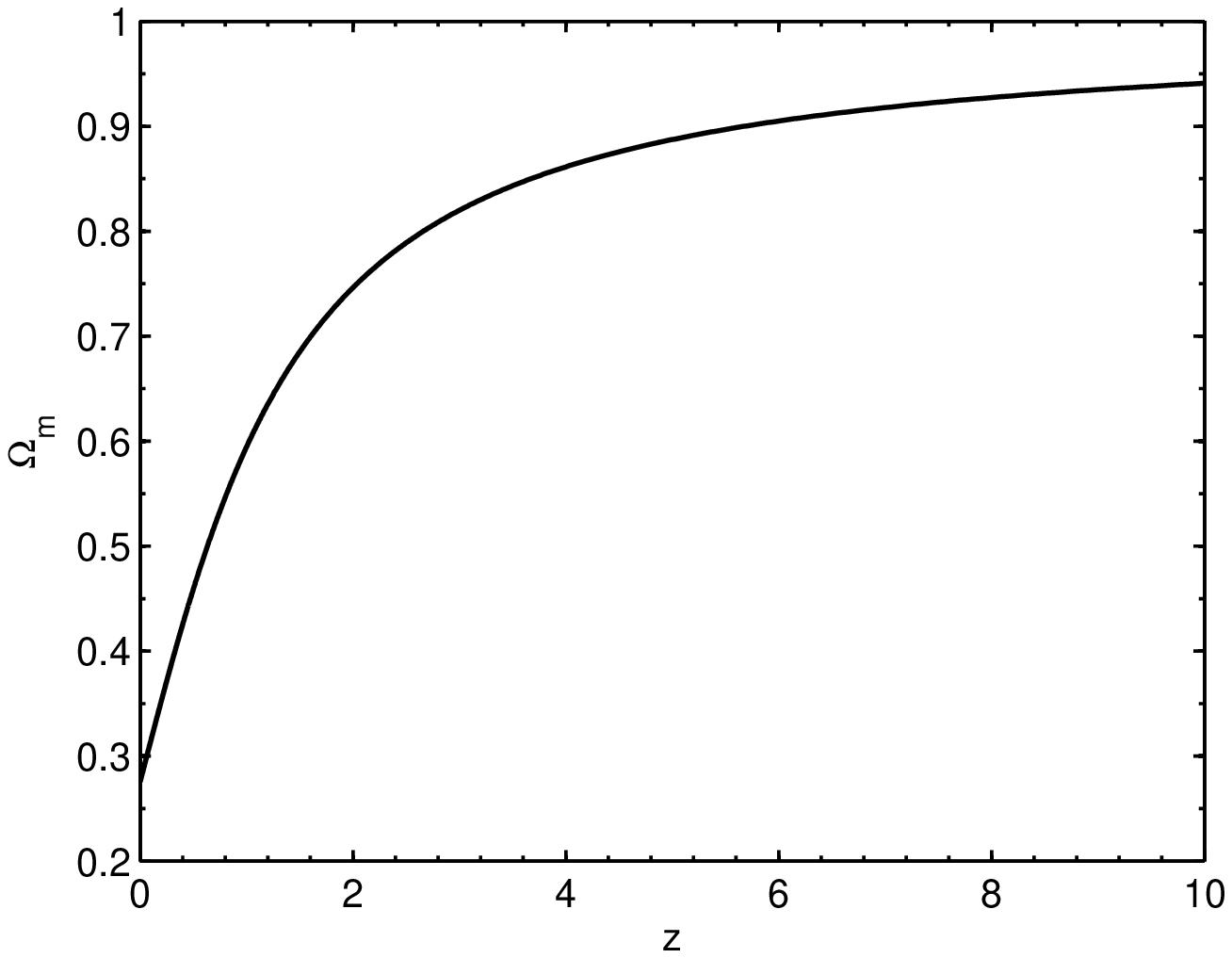} 
\includegraphics[width=0.4\linewidth]{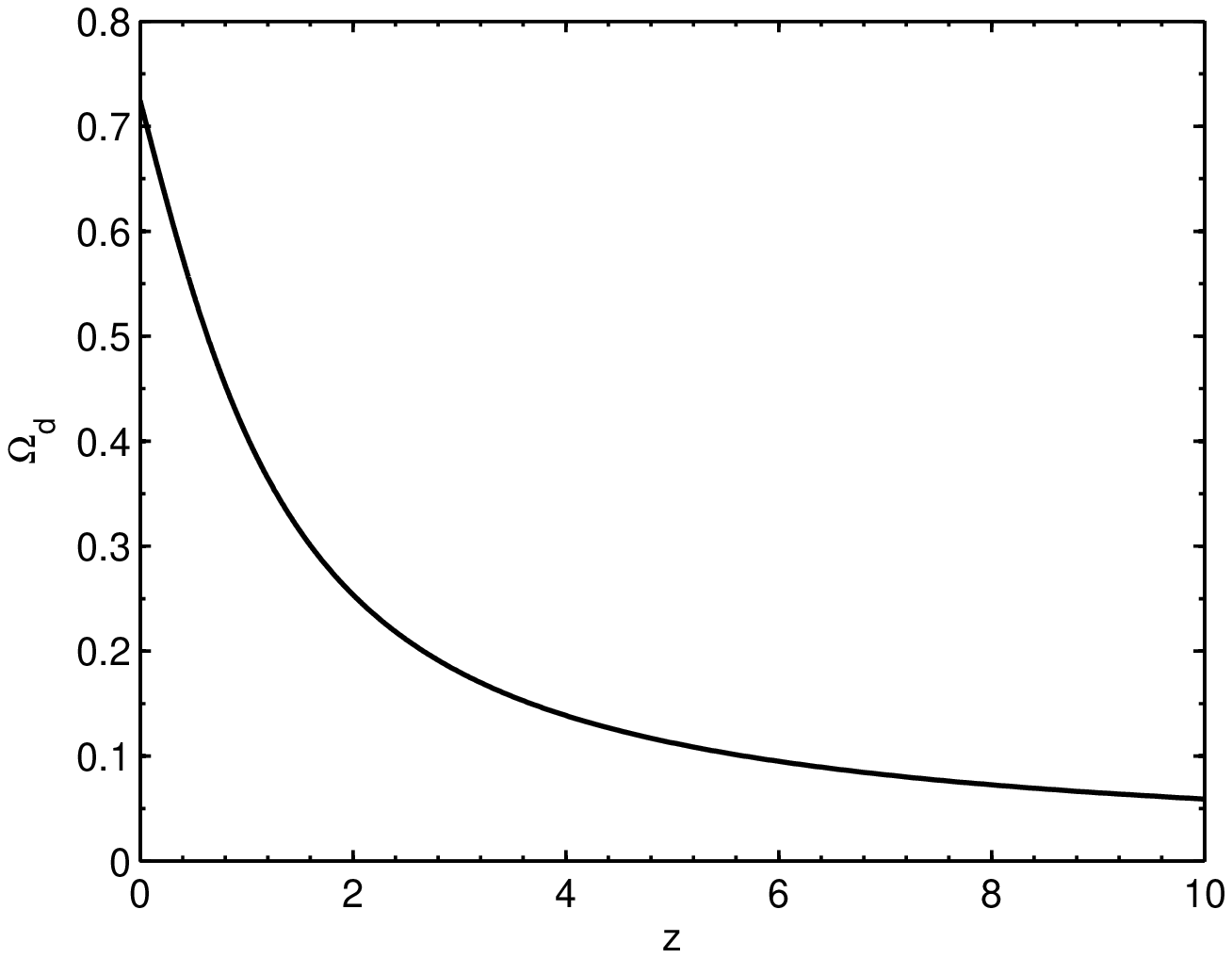}
\caption{The evolution of EoS parameter $w_d$ (left-top panel),
deceleration parameter $q$ (right-top panel), mater density parameter $\Omega_{\rm m}$
(left-bottom panel) and dark energy density parameter $\Omega_{\rm d}$ (right-bottom panel) for the best fit values in the 
$G$-corrected HDE model. 
}\label{fig:EoS}
\end{figure}

Then we calculate the evolution of some cosmological quantities: EoS
parameter of the dark energy component $w_{\rm d}$, matter and dark energy density parameters, and deceleration parameter
for the G-corrected HDE model based on the best fit values of
cosmological parameters in table \ref{tab:MCMC}.
In the top-row of figure \ref{fig:EoS}, the evolution of the  EoS parameter $w_d$ (left
panel) and the deceleration parameter $q$ (right panel) in terms of
the redshift parameter $z$ has been plotted by solving equations
(\ref{OL}) and (\ref{18}) and using (\ref{12}).  We see that  by using the best fit values in the 
G-corrected HDE model, within $1 \sigma$ confidence level, one  obtains the present
value of EoS parameter as: $−1.887<w_{\rm d0}<-0.232$ 
which can enters to the phantom regime in lower bound. It is worthwhile
to mention that in this case the phantom regime can be achieved without
invoking interaction between dark matter and dark energy. In the left
panel, the parameter $q$ can transit from positive values $q>0$ to
negative values ($q<0$) which indicates the transition from early
decelerated expansion to current accelerated phase of expansion. The
present value of the deceleration parameter $q$ within $1 \sigma$ confidence level is obtained as: $−1.1268<q_{0}<−0.5565$. 
Finally, the evolution of density parameters of dark
energy and pressure-less matter  has been shown  in the bottom row figure
\ref{fig:EoS}. The density parameter of the 
pressureless matter decreases and dark energy increases by decreasing redshift, indicating
the early time CDM dominated universe and current dark energy
dominated phase in G-corrected HDE cosmology.

\section{conclusion}\label{sec:conc}
We performed cosmological constrains  on the parameters
of the holographic dark energy model with time varying gravitational coupling $G$ using a
Markov chain Monte Carlo simulation. We used the SNIa, CMB, BAO
and X-ray mass gas fraction data for data fitting. In the framework of the  modified
Friedmann equations, we obtained the best fit values for the
cosmological parameters as: the physical baryon matter density
$\Omega_{\rm b}h^2=0.0222^{+0.0018+0.0021}_{-0.0013-0.0016}$, dark
matter physical density $\Omega_{\rm
c}h^2=0.1121^{+0.0110+0.0130}_{-0.0079-0.0096}$, Hubble parameter at
the current time $H_0=69.8809^{+3.5339+4.1638}_{-3.4423-4.4567}$ and
the age of the Universe $13.8094^{+0.2801+0.3776}_{-0.3618-0.4392}$.
We constrained the G-corrected HDE parameters $c$ and $\alpha_{\rm G}$ as
well. The best fit value of the  parameter $c=0.9322^{+0.4569}_{-0.5447}$
is in agreement with results of the previous works
\cite{Zhang:2013mca, Zhai:2011pp, Lu:2009iv}.
In our model the best fit value
for the rate of changing the gravitational coupling constant with time is
$\dot{G}/G=+1.14\times10^{-11}{\rm yr^{-1}}$. This value is close to
the value obtained by others like constraints in
\cite{Benvenuto:2004bs, Biesiada:2003sr}.
Therefore the result of our analysis is compatible with observations
and other analysis of the  HDE model and time varying
gravitational coupling constant.
\par The evolution of the 
deceleration parameter $q$, for the best fit values of cosmological
parameters, indicates the transition from past decelerated to
current accelerated expansion. By using the best fit values of the aforementioned parameters, 
within $1\sigma$ CL, the
phantom regime $w<-1$ can be achieved in this model.

\par In summary we conclude that the holographic dark energy with a time varying
gravitational coupling constant in the framework of the modified Friedmann equations, could be
a candidate to describe the accelerated expansion of the universe. In addition, in future works,  by using the data
from Planck \cite{Ade:2013zuv} and nine-year WMAP \cite{Bennett:2012zja} projects, one can make the constraints on
the model parameters even tighter.

\section*{Acknowledgements} 
H. Alavirad would like to thank J. M. Weller for helpful and useful discussions 
and comments. 
\section*{References}

\bibliographystyle{utphys}
\bibliography{references}

\end{document}